# Info-Computationalism and Philosophical Aspects of Research in Information Sciences

Gordana Dodig-Crnkovic
Mälardalen University, Sweden,
http://www.idt.mdh.se/personal/gdc

19th November, 2008

The historical development has lead to the decay of Natural Philosophy which until 19th century included all of our knowledge about the physical world into the growing multitude of specialized sciences, within "The Classical Model of Science". The focus on the in-depth enquiry disentangled from its broad context lead to the problem of loss of common world-view and impossibility of communication between specialist research fields because of different languages they developed in isolation. The need for a new unifying framework is becoming increasingly apparent with the information technology enabling and intensifying the communication between different research fields, knowledge communities and information sources. This time, not only natural sciences, but also all of human knowledge is being integrated in a global network such as Internet with its diverse knowledge and language communities.

Info-computationalism (ICON) as a synthesis of pancomputationalism and paninformationalism presents a unifying framework for understanding of natural phenomena including living beings and their cognition, their ways of processing information and producing knowledge. Within ICON physical universe is understood as a network of computational processes on an informational structure. The matter/energy in this model is replaced by information/computation where information is the structure, whose dynamics are identified as natural computation.

ICON is an example of philosophical framework in a direct connection with the related scientific fields, and the process is mutual exchange: scientific findings influence philosophical thinking and vice versa. Research is going on in Natural computing on modeling natural phenomena including living organisms as info-computational agents and implementing natural computation principles into technological artifacts. Lessons learned from the design and implementation of our understanding of living natural computational agents leads to artifacts increasingly capable of simulating essential abilities of living organisms to process and structure information. Among others, ICON supports scientific understanding of the mind (perception, thinking, reasoning, will, feelings, memory, etc.) providing computational naturalist models of cognition.

1. **Science and Philosophy**

"Our best efforts are directed at finding out why things are as they are or why the events around us occur as they do. This is so in all disciplines and philosophy is no exception in this regard; what sets philosophy apart is that it probes deeper as well as being more general. It queries the presuppositions other disciplines leave untouched, and in trying to clarify such presuppositions and trace their interconnections it seeks to find out how the world is put together and how it works." (Pivčević, 2007)

Knowledge, both propositional and non-propositional is the basic constituent of science produced by research process. Nevertheless, often knowledge is considered as identical with propositional knowledge and science is identified with a search for *truth* about the world; *truth* considered to be in a propositional form. Though History, Sociology and Philosophy of Science all offer good reasons for seeing science as a goal-driven human activity aimed at production of *models* which enable us to predict, correlate and structure (compress, according to Chaitin) relevant information about the world.

The role of science as information compression mechanism (in a sense of Chaitin's algorithmic information, Chaitin 2007) is especially visible in its search for simple and universal laws, which especially characterize physical sciences. Historically however the idea of natural law has significantly evolved. The ancient idea of *deterministic order* of the universe was closely related to the *principle of causality* that denotes a necessary relationship between one event (cause) and another event (effect) which is the direct consequence (result) of the first. However, *indeterminism* of quantum physics induced new elements into the picture of natural laws. Later on *disorder* was found even in rule-governed systems, showing that deterministic functions can generate unpredictable results.

What is crucial to the scientific knowledge is not its *certainty*, otherwise not much would qualify as knowledge in the history of science. Even understanding of fundamental scientific ideas such as time, space, mass, and trajectory have been successively historically revised. What has constantly been characterizing sciences is the *rationality* of their approaches, presuppositions and aims. Science is primarily an explanatory and predictive *tool of making sense* (Pivčević).

Nowadays, complexity of real-world problems becomes a focus of sciences, in the first place thanks to computational capabilities of ICT. Instead of static, symmetric and steady-state, eternal static order, a new dynamic picture emerges in which everything changes, and order is created as a pattern over layers of pulsating underlying physical processes, and very simple basic rules can lead to evolution of complex systems.

**2. The Classical Model of Science and the Complexity of the Real World**

When analyzing the relationship between Philosophy and Science it is important to recognize the paradigm shift going on in both sciences and philosophy as a consequence of recent growth of trans-disciplinary, interdisciplinary and cross-disciplinary research unfamiliar to classical philosophical analysis. Research practices and the resulting sciences are today not what they used to be before ICT revolution. Effective information processing, storage and exchange capabilities of today's networked global research communities make possible addressing of complex problems which were impracticable before because of their informational and computational complexity. Such problems are typically defined at several levels of abstraction (levels of description, levels of functionality) and thus they usually cover several classical research disciplines.

This development towards complex problems made some presuppositions of existing sciences explicit. Among others, their domains of validity have become visible, making distinctions and relationships between different fields clearer and easier to recognize. Typical example where domain-dependence of different sciences becomes apparent is in modeling of living organisms.

At the basis of our common intuitive understanding there is an idealized view of sciences, The Classical Model of Science, which according to (Betti & De Jong, 2008) is a system S of propositions and concepts satisfying the following conditions:

> "(1) All propositions and all concepts (or terms) of S concern a specific set of objects or are about a certain domain of being(s).
> 
> (2a) There are in *S* a number of so-called *fundamental concepts* (or terms).
> 
> (2b) All other concepts (or terms) occurring in *S* are *composed of* (or are *definable from*) these fundamental concepts (or terms).
> 
> (3a) There are in *S* a number of so-called *fundamental propositions*.
> 
> (3b) All other propositions of *S follow from* or *are grounded in* (or *are provable* or *demonstrable from*) these fundamental propositions.
> 
> (4) All propositions of *S* are *true*.
> 
> (5) All propositions of *S* are *universal* and *necessary* in some sense or another.
> 
> (6) All propositions of *S* are *known to be true*. A non-fundamental proposition is known to be true through its *proof* in *S*.
> 
> (7) All concepts or terms of *S* are *adequately known*. A non-fundamental concept is adequately known through its composition (or definition).

The Classical Model of Science is a reconstruction a posteriori and sums up the historical philosopher's ideal of scientific explanation. (Betti & De Jong, 2008)

The fundamental is that "All propositions and all concepts (or terms) of S concern *a specific set of objects or are about a certain domain of being(s)*."

This view of science together with exponential growth of scientific knowledge has lead to the extreme compartmentalization and specialization of sciences which can work well in some cases, while in others its aim and meaning may be questioned. One example might be medicine with its many narrow specializations where a patient is treated by different specialists who see her/his health in only one specific domain and administer medicines as if the patient would be a sum of disjoint domains and not one single organism in whom all different domains overlap and interact. This compartmentalization has its historical roots in the limitations of human information processing capabilities. Today's info-computational networks present good tools for information processing and exchange and they enable building of complex knowledge structures in which experts with different specialties can adaptively interact and make well informed decisions taking into account existing knowledge from other fields.

ICT is already affecting the way research is done and scientific knowledge is produced which will make "The Classical Science" just an element of a complex knowledge production structure.

After the ages of idealizations and compartmentalization, sciences are starting to examine their own roots, presuppositions and mutual relationships (contexts awareness). Computer-aided science resembles electrified industry – a new world of possibilities presents itself compared to the pre-ICT world. With nowadays information and knowledge management technologies sciences can afford to take into account and model complexity of the real world with huge variety of parameters, complex structures and intricate dynamics. Instead of taking idealized frozen slices of reality, we can make realistic models, simulate and study the dynamics of complex systems and their interactions. The step from the "old" "pre-computational" world to the new ICT-world is not a trivial one. Often conceptual confusions arise about different kinds of knowledge, the domain of its applicability, the underlying presuppositions, the relationships with other possible knowledge and similar.

## 3. Info-Computationalism as a Framework for Unity of Knowledge

Info-computationalism is a view according to which the physical universe on a fundamental level can be understood as an informational structure which dynamics is computational process. The matter/energy in this model is replaced by information/computation; matter (structure) corresponds to information while the dynamics - constant changes in the informational structure – are computational processes. In this view the universe is a huge computer network which by physical laws "computes" its own next state (Chaitin, pancomputationalism). Information is the fabric of the universe. An instantaneous "snap" of the universe reveals the structure. Changes are computational processes. Computation is simply information processing.

ICON unites pan-computationalism (natural computationalism) with pan-informationalism. In short, the Info-Computationalism is a dual-aspect approach based on two fundamental concepts: information and computation. On this view, as many computationalists already have declared (Zuse, Fredkin, Wolfram, Chaitin, Lloyd, ..) the universe is a huge computing system (or a network of computing processes) which by means of physical laws compute its own next state. *It must be pointed out that the computing universe is not identical with (or reducible to) today's computers.* Rather, computing is what the whole of the universe does while processing its own information by simply following natural laws. ("Natural computing" in its different forms. [How much of "natural laws" are orderly and how much is random is a separate discussion, see Chaitin's work].

One might suspect that the computationalist idea is vacuous, and if everything is info-computational, then it says nothing about the world. The computationalist claim however should be understood as similar to the claim that universe is made of atoms. Atom is a very useful concept which helps understanding the world in many fields. So is the info-computational approach. Universe is NOT "nothing but atoms", but on some view (level of organization, level of abstraction) may be seen as atoms.

Unlike many other (pan)computationalists I do not presuppose that computationalism necessarily implies digital computing. As Seth Lloyd points out, on the basic quantum-mechanical level discrete and analogue, digital and continuous computing is going on. See more about the question of digital/analog universe in (Dodig-Crnkovic, 2006)

As already emphasized, physical reality can be addressed at many different levels of organization. Life and intelligence are the phenomena especially characterized by info-computational structures and processes. Living systems have the ability to act autonomously and store information, retrieve information (remember), anticipate future behavior in the environment with help of information stored (learn), adapt to the environment (in order to survive). In Epistemology Naturalized (Dodig-Crnkovic, 2007) I present a model which helps understanding mechanisms of information processing and knowledge generation in an organism. Thinking of us and the universe as a network of computational devices/processes allows easier approach of question about boundaries between living and non-living beings.

Info-computationalism sees our bodies as advanced computational machines in constant interaction with the environmental computational processes. Our brains are informational architectures undergoing computational processes on many levels of organization. On the levels of basic physical laws there is a computation going on. All which physics can conceptualize, describe, calculate, simulate and predict can be expressed in info-computationalist terms. On the level of molecules (with atoms and elementary particles as structural elements) there are computational processes going on. The nerve cell level can be understood as the next level of relevance in our understanding of the computational nature of the brain processes. Neurons are organized in networks, and with neurons as building blocks new computational phenomena appear on the level of neural network. The intricate architecture of informational structures in the brain, implementing dif-ferent levels of control mechanisms not unlike virtual machines on higher level running on the structure below. What we call "informational architecture" is fluid and interactive, not so much crystal-lattice-type rigid construction but more like networks of agents.

The development is going on in two directions: analyzing living organisms as info-computational systems/agents, and implementing natural computation strategies (organic computing, bio computing) into artifacts. Lessons learned from the design and implementation of our understanding of living natural computational agents through iterative process of improvements will lead to artifacts that in increasingly higher degree will be capable of simulating characteristics of living organisms.

**4.   Information & Computing Sciences and Philosophy**

When discussing the relationship between Philosophy and Sciences, an instructive example can be found in Computing and Philosophy (CAP) field. The following is the list over some of research fields presented at CAP conferences: *Philosophy of information, Philosophy of computation, Computational approaches to the problem of mind, Philosophy of Computing, Real and virtual, modeling, simulations, emulations, Computing and Information Ethics, Societal aspects of computing and IT, Philosophy of Complexity, Computational metaphysics, Computational Epistemology, Computer-based Learning and Teaching*. From the list it is evident that CAP represents a forum for cross-disciplinary – inter-disciplinary – multi-disciplinary knowledge exchange and establishment of relationships between existing knowledge fields, and philosophical reflection over them. This development of a new body of knowledge is followed by a distinct paradigm shift in the knowledge production mechanisms. (Dodig-Crnkovic, 2003) Globalization, information networking, pluralism and diversity expressed in the cross-disciplinary research in a complex web of worldwide knowledge generation are phenomena that need to be addressed on a high level of abstraction, which is offered by philosophical discourse. Examples of philosophical approaches closely connected to the on-going paradigm shift may

be found in (Floridi, 2004, 2005; Wolfram 2003, Mainzer 2003 and 2004, Chaitin 2005, Lloyd 2006, Zuse 1967).

In order to understand various important facets of ongoing info-computational turn and to be able to develop knowledge and technologies, a dialogue and research on different aspects of computational and informational phenomena are central. Taking information as a fundamental structure and computation as information processing (information dynamics) one can see the two as complementary, mutually defining phenomena. No information is possible without computation (information dynamics), and no computation without information. (Dodig-Crnkovic 2005, Dodig-Crnkovic & Stuart 2007)

## 5. Knowledge as a Complex Informational Architecture. The Necessity of a Multidisciplinary Dialogue

Why is it important to develop a multi-disciplinary discourse which will present the departure from the monolithic "Classical Model of Science" caused by diversity of the domains, methods and levels of organization/levels of abstraction? The main reason is epistemological – multidisciplinarity provides the fundamental framework suitable for common understanding and communication between presently disparate fields. This argument builds on a view of knowledge as informational construction. According to Stonier (1997), data is a series of disconnected facts and observations, which is converted into information by analyzing, cross-referring, selecting, sorting and summarizing the data. Patterns of information, in turn, can be worked up into knowledge which consists of an organized body of information. Stonier's constructivist view emphasizes two important facts:

— going from data to information to knowledge involves, at each step, an input of work, and

— at each step, this input of work leads to an increase in organization, thereby producing a hierarchy of organization.

Research into complex phenomena (Mainzer 2004) has led to an insight that research problems have many different facets which may be approached differently at different levels of abstraction and that every knowledge field has a specific domain of validity.

This new understanding of a multidimensional many-layered knowledge space of phenomena have among others resulted in an vision of an ecumenical conclusion of science wars by recognition of the necessity of an inclusive and complex knowledge architecture which recognizes importance of a variety of approaches and types of knowledge. [See for example Smith and Jenks, 2006.] Based on sources in philosophy, sociology, complexity theory, systems theory, cognitive science, evolutionary biology and fuzzy logic, Smith and Jenks present a new interdisciplinary perspective on the *self-organizing complex structures*. They analyze the relationship between the process of self-organization and its environment/ecology. Two central factors are the role of information in the building of complex structures and the development of topologies of possible outcome spaces. The authors argue for a continuous development from emergent complex orders in physical systems to cognitive capacities of living organisms, complex structures of human thought and to cultures. This is a new understanding of unity of interdisciplinary knowledge, unity in structured diversity, also found in (Mainzer 2004).

In a complex informational architecture of knowledge, logic, mathematics, quantum mechanics, thermodynamics, chaos theory, cosmology, complexity, the origin of life, evolution, cognition, adaptive systems, intelligence, consciousness, societies of minds and their production of knowledge and other artifacts … there are two basic phenomena in common: *information and computation* which provide a framework for those jigsaw puzzle pieces of knowledge to be put together into a complex and dynamic unified info-computational view.

The body of knowledge and practices in computing and information sciences, as a new research field, has grown around an artifact – a computer. Unlike old research disciplines, especially physics which has deep historical roots in Natural

Philosophy, research tradition within computing community up to now was primarily focused on problem solving and has not developed very strong bonds with philosophy[1]. The discovery of philosophical significance of computing in both philosophy and computing communities has led to a variety of new and interesting insights on both sides.

The view that information is the central idea of Computing/Informatics is both scientifically and sociologically indicative. Scientifically, it suggests a view of Informatics as a generalization of information theory that is concerned not only with the transmission/communication of information but also with its transformation and interpretation. Sociologically, it suggests a parallel between the industrial revolution, which is concerned with the utilizing of energy, and the information revolution, which is concerned with the utilizing of information. (Dodig-Crnkovic, 2003)

**6. The Relevance of Philosophy for Sciences and Sciences for Philosophy**

The development of Philosophy is sometimes understood as its defining new research fields and then leaving them to sciences for further investigations (Floridi's lecture in Swedish National PI course on the development of Philosophy), (PI, 2004). At the same time, Philosophy traditionally also learns from sciences and technologies, using them as tools for production of the most reliable knowledge about the factual state of affairs of the world. We can mention a fresh example of current progress in modeling and simulation of brain and cognition that is of vital importance for the philosophy of mind. As so many times in history, the first best approach when scarce empirical knowledge exists, the intuitive one does not necessarily need to be the best. Wolpert (1993) for example points out that science is an *unnatural mode of thought*, and it very often produces a counterintuitive knowledge, originating from the experiences with the world made by tools different from everyday ones, experiences in micro-cosmos, macro-cosmos, and other areas hidden for our unaided cognition.

---

[1] Alan Turing was one of the notable exemptions to the rule. Others are Weizenbaum, Winograd and Flores (G A Lanzarone, 2007). It should also be mentioned that Computing always had strong bonds with logic, and that especially AI always had recognized philosophical aspects.

A good example of "unnatural" character of scientific knowledge is a totally counterintuitive finding of astronomy that Earth is revolving around the Sun and not the vice versa as our intuitions would tell us. At present, similar Copernican Revolution seem to be going on in the Philosophy of Mind, Epistemology (understood in informational terms), in Philosophy of Information, and Philosophy of Computing. The recently published book, 'Every Thing Must Go: Metaphysics Naturalised', (Ladyman at al. 2007) rightly argues for Philosophy (and specifically Mataphysics) informed by the latest developments in special sciences, instead of philosophers' a priori intuitions and common sense. Specialist sciences such and Cognitive Science and Neuroscience e.g. have collected valuable knowledge that should be adopted by Philosophy of Mind instead of relying on historical ideas based on common sense. The process going on in the opposite direction, from philosophy to specialist sciences has been mentioned several times in this article in example of Info-Computationalism. Additionally, Ethics must be mentioned as a branch of Philosophy of increasing importance for sciences.

Sciences (Wissenschaften) are nowadays understood as rational tools for information compression which enable us to predict and efficiently handle different phenomena, and not (as previously believed) a source of *absolute truth* about the universe revealing the divine plan while reading the "book of nature". Instead, they are telling us at least as much about who we are in the relationship with the world – in that new light a conscious reflection over our values, motives, priorities and ways is vital. Especially when it comes to applied science and technology, the relevance of ethical analysis is obvious. Computers as our primary tools of information processing in research and otherwise can be used for good or bad and newly developed technologies can be misused. The challenge of radically new applied sciences and technologies is in what James Moor (1985) calls "policy vacuums" – it is not often possible to predict in what ways technology can be misused and what end results it might have. Ethical judgment is absolutely necessary and in the long run, even in the education system, research ethics, computing ethics, information ethic and other specific branches of applied ethics accompanying specialized sciences and technologies should become an integral part of education and research system.

**7.   Answering some Usual Kinds of Criticism of Info-Computationalism**

Info-Computationalist view may be interpreted as a claim that the whole of the world is "nothing but a (computational) machine" and that we humans are essentially machines with no free will or feelings. That is obviously not the case. The view that the universe is an info-computational phenomenon means that the universe as it is may be understood and modeled in info-computational language. Feelings, qualia and other mental phenomena are emergent properties of the physical world which is info-computational.

The role of different paradigms in our understanding of the universe can be analyzed on historical examples. In the past, several major paradigm shifts occurred: from mytho-poetic to mechanistic to the emerging computationalist understanding of the universe. Consequently, we can ask the same question about mytho-poetic and mechanist universe: Was that understanding of the universe true? Was it real or merely metaphoric? For the mytho-poetic universe, the answer is simple – it was a metaphor. Even though mechanicism was primarily the outlook at the inanimate matter, and mechanistic approaches to robotics (mechanical quasi-humans) did not work for any other purpose but the entertainment, mechanistic worldview nevertheless helped us to learn a lot about the universe, in the first place inanimate but even many basic facts about living world (such as that *there is no Élan vital* but the same physical mechanistic laws govern the whole of the physical universe, living as well as non-living).

The parallel development goes on in the course of computationalism now again. We will learn about informational and computational resources and capabilities, of the universe and we will develop even more powerful ways of learning, via intelligent systems we will successively improve.

Knowing that biological organisms (including humans) are information-processing "machines" does not make them less fascinating. In the same way as knowing that all of us are made of atoms does not mean that we do not have free will, imagination and real feelings. Understanding fundamental level processes does not make music, arts and philosophy obsolete.

Info-computationalism helps us both by supplying the tools for knowledge and artifact production and even tools for understanding of the natural phenomena and artifacts on many different levels. That is also why philosophy is coming back to sciences based on info-computational knowledge. A holistic, high level of abstraction view is necessary as a self-reflective process of knowledge.

In sum: Info-Computationalism is by no means the *final answer* to 'Life!' 'The Universe!' and 'Everything!' (which is 42, as we learned from Adams as quoted by Vincent Müller in this volume), but a learning tool which will help us again reach a unity of knowledge on one specific level of abstraction: info-computational.
So the answer (Info-Computationalism) is not *The final answer*, but all the same seems to be a reasonable answer to a reasonable question: how to get common language for disparate specialist fields which can enable mutual understanding and building of new knowledge?

**References**

Betti A & De Jong W. R., (guest edit), The Classical Model of Science I: A Millennia-Old Model of Scientific Rationality, Forthcoming in Synthese, Special Issue https://www.surfgroepen.nl/sites/cmsone/cmsoneweb/index.html

Chaitin G.J., Epistemology as Information Theory. Alan Turing Lecture given at E-CAP 2005. In: Dodig-Crnkovic G. and Stuart S., eds. Computation, Information, Cognition – The Nexus and The Liminal, Cambridge Scholars Publishing, Cambridge UK (2007) Available at
http://www.cs.auckland.ac.nz/CDMTCS/chaitin/ecap.html

Dodig-Crnkovic G., Shifting the Paradigm of the Philosophy of Science: the Philosophy of Information and a New Renaissance, Minds and Machines,

2003, Volume 13, 4 Available at:
http://www.springerlink.com/content/g14t483510156726
http://www.idt.mdh.se/personal/gdc/work/shifting_paradigm_singlespace.pdf

Dodig-Crnkovic G., Investigations into Information Semantics and Ethics of Computing, Mälardalen University (2006) Available at: http://www.idt.mdh.se/personal/gdc/work/publications.html

Dodig-Crnkovic G. and Stuart S., eds. Computation, Information, Cognition – The Nexus and The Liminal, Cambridge Scholars Publishing, Cambridge UK (2007), Available at: http://www.idt.mdh.se/ECAP-2005/Intro-Preface-ComputationInformationCognition.pdf

Dodig-Crnkovic G., Epistemology Naturalized: The Info-Computationalist Approach, APA Newsletter on Philosophy and Computers, Spring 2007 Volume 06, Number 2 (2007) Available at: http://www.apa.udel.edu/apa/publications/newsletters/v06n2/Computers/04.asp

Floridi L., Open Problems in the Philosophy of Information, Metaphilosophy, 35.4 (2004)

Floridi L. (ed), (2004) The Blackwell Guide to the Philosophy of Computing and Information (Blackwell Philosophy Guides),

Fredkin E, Digital Philosophy (and references therein) Available at : http://www.digitalphilosophy.org/Home/Papers/tabid/61/Default.aspx

Ladyman J and Ross D, with David Spurrett and John Collier, Every Thing Must Go: Metaphysics Naturalized. Oxford University Press 2007

Lanzarone G A, In Search of a New Agenda, APA Newsletter Computing and Philosophy, Fall 2007

Lloyd, S. Programming the Universe: A Quantum Computer Scientist Takes on the Cosmos. Jonathan Cape (2006)

Mainzer, K. Thinking in complexity, Springer, Berlin, 2004

Mainzer, K. Computerphilosophie zur Einführung , Junius, Hamburg, 2003

Moor, J. What Is Computer Ethics? Metaphilosophy 16/4 (1985): 266-275.

Pivčević E., The Reason Why: A Theory of Philosophical Explanation, KruZak, (2007)

PI, Swedish National Course, http://www.idt.mdh.se/~gdc/PI_04/index.html (2004)

Smith, J. and Jenks C., Qualitative Complexity Ecology, Cognitive Processes and the Re-Emergence of Structures in Post-Humanist Social Theory, Routledge, 2006.

Stonier T. (1993) The Wealth of Information, Thames/Methuen, London

Stonier T. (1997) Information and Meaning. An Evolutionary Perspective, Springer, Berlin, New York

Wolfram S., A New Kind of Science, Wolfram Science (2002)

Wolpert L., The unnatural nature of science, Harvard University Press (1993)

Zuse, K., Rechnender Raum. Elektronische Datenverarbeitung, vol. 8, 336--344 (1967)